\documentclass{elsart}

\usepackage{graphicx}
\usepackage{amssymb}
\begin{document}
\begin{frontmatter}


\title{Mobile Localization in Nonlinear Schr\"odinger Lattices}
\author{J. G\'omez-Garde\~nes}\ead{gardenes@unizar.es}, 
\author{F. Falo} and 
\author{L.M. Flor\'{\i}a}
\address{Dpt. de F\'{\i}sica de la Materia Condensada and Instituto de
Biocomputaci\'on y F\'{\i}sica de los Sistemas Complejos (BIFI), Universidad
de Zaragoza, 50009 Zaragoza, Spain.}
\address{Dept. de Teor\'{\i}a y Simulaci\'on de Sistemas Complejos,
Instituto de Ciencia de Materiales de Arag\'on (ICMA), C.S.I.C.-Universidad
de Zaragoza, 50009 Zaragoza, Spain.}
\begin{abstract}
Using continuation methods from the integrable 
Ablowitz-Ladik lattice, we have studied the structure 
of numerically exact mobile discrete breathers in the 
standard Discrete Nonlinear Schr\"odinger equation. 
We show that, away from that integrable limit, the mobile 
pulse is dressed by a background of resonant plane waves 
with wavevectors given by a certain selection rule. 
This background is seen to be essential for supporting mobile
localization in the absence of integrability. We show how the variations of the
localized pulse energy during its motion are balanced by the
interaction with this background, allowing the localization 
mobility along the lattice.  
\end{abstract}

\begin{keyword}
Nonlinear dynamics, Localized modes 
\PACS 05.45.-a; 63.20.Pw 
\end{keyword}
\end{frontmatter}

\section{Introduction}
\label{s.intro}
The phenomenon of intrinsic localization (collapse 
to self-localized states) due to nonlinearity in 
discrete systems governed by Schr\"odinger equations 
is of fundamental interest in Nonlinear Physics 
\cite{Scott,Cai}, and is the subject of current 
active experimental research in several areas 
like nonlinear optics \cite{Christo}, Bose-Einstein 
condensate arrays \cite{Strecker,Cataliotti,Smerzi}, 
polaronic effects in biomolecular processes, and 
local (stretching) modes in molecules and molecular 
crystals (see \cite{Scott,Eilbeck,Chaos} and 
references therein).
Discrete Non-Linear Schr\"odinger equations (NLS 
lattices for short) provide the theoretical description 
of these systems, where pulse-like (self-localized) 
states are observed. 

The standard Discrete Nonlinear 
Schr\"odinger (DNLS) equation is the (simplest) 
discretization of the one-dimensional continuous 
Schr\"odinger equation with cubic nonlinearity in 
the interaction term, {\em i.e.}, 
\begin{equation}
{\mbox i} \dot{\Phi}_n= -(\Phi_{n+1} + \Phi_{n-1}) - \gamma |\Phi_n|^2 \Phi_n\;,
\label{DNLS}
\end{equation}
where $\Phi_n(t)$ is a complex function of time. The first term 
on the right takes account of the dispersion
and the second of the nonlinearity, the parameter $\gamma$ 
is the ratio between them. For the Bose-Einstein 
condensate lattices dealt with in 
\cite{Strecker,Cataliotti,Smerzi} one can think of 
$\Phi_n$ as the boson condensate wavefunction in the 
$n$-th (optical) potential well, and $\gamma$ would 
thus be related to the so-called $s$-wave scattering 
length \cite{Leggett}. The self-focussing effect of local nonlinearity 
balanced by the opposite effect of the dispersive 
coupling makes possible the existence of localized 
boson states in the Schr\"odinger 
representation of the condensate lattice (Gross-Pitaevskii equation). 
In a localized state (discrete 
breather) of the boson lattice the profile of 
$|\Phi_n|^2$ decays exponentially away from the 
localization center. These solutions have an internal frequency, 
$\Phi_n=|\Phi_n|\exp({\mbox i}\omega_b t)$, so that the discreteness
is essential to avoid resonances with the phonon band and keeping
localized the energy. Pinned (immobile) localized solutions of eq. 
(\ref{DNLS}) have been rigorously characterized 
\cite{MackayAubry} and extensively studied by 
highly accurate numerical \cite{Johansson} and 
analytical approximations. However, for exact {\em mobile} 
discrete breathers no rigorous formal proof of existence in standard 
DNLS is available nowadays although lot of works have studied these
kind of solutions (see e.g. \cite{CretegnyAubry,Duncan,Kladko,Musslimani}). 

The translational motion of discrete breathers introduces a new time 
scale (the inverse velocity) into play, so generically a moving
breather should excite resonances with the plane wave band expectra. 
In a hamiltonian system, these radiative losses would tend to
delocalize energy and some compensating mechanism is needed in order 
to sustain exact stationary states of breather translational motion. 
To address the problem we use unbiased ({\em i.e.} not based on
ansatze on the expected functional form of the exact solution) and 
precise numerical methods which allow observations of numerically 
exact non-integrable mobility, paving the way to further physical 
(and mathematical) insights.

In this letter, after explaining in section \ref{s.model} the basis of the numerical method 
(fixed point continuation from the integrable 
Ablowitz-Ladik limit \cite{AL}) and its relevant 
technical details briefly, we will discuss the 
structure of the discrete NLS breathers in \ref{s.mdb}. They are 
found to be the exact superposition of a travelling 
exponentially localized oscillation (the {\em core}), 
and an extended ``background'' built up of finite 
amplitude plane waves $A \exp[{\mbox i}(k n-\omega t)]$. 
These resonant plane waves fit well simple (thermodynamic limit)
predictions based on discrete  symmetry requirements. Finally in
section \ref{s.bckg} we show how the resonant
background is seen to be an indispensable part of the solution. In
this regard we present the mechanism through which the interaction
core-background compensates the variations of the core energy 
(no longer an invariant of motion away from the integrable limit), 
during the translational motion.

\section{Salerno Model and Continuation Method}
\label{s.model}
The method used here makes use of the following NLS lattice, 
originally introduced by Salerno \cite{Salerno}, 
\begin{equation}
{\mbox i} \dot{\Phi}_n= -(\Phi_{n+1} + \Phi_{n-1})\left[ 1 + \mu |\Phi_n|^2 \right] 
- 2 \nu \Phi_n |\Phi_n|^2\;.
\label{Salerno}
\end{equation}
This lattice, though non-integrable for $\nu \neq 0$, provides 
a Hamiltonian interpolation between the standard DNLS 
equation (\ref{DNLS}), for $\mu = 0$ and $\nu = \gamma/2$, 
and the integrable Ablowitz-Ladik lattice \cite{AL}, A-L for short, 
when $\mu = \gamma/2$ and $\nu = 0$. The A-L model is a remarkable integrable lattice 
possessing a family of exact moving breather solutions:
\begin{eqnarray}
\Phi_n (t) &=&\sqrt{\frac{2}{\gamma}} \sinh \beta \; \mbox{sech}
[\beta (n-x_0(t))] \exp [{\mbox i}(\alpha (n-x_0(t)) +\Omega(t))],
\label{A-Lbreather}
\end{eqnarray}
the two parameters $\omega_b$ and $v_b$ are the breather frequency 
and velocity 
\begin{equation}
\omega_b\equiv\dot{\Omega}(t)=2\cosh \beta \; \cos \alpha \;+\; \alpha v_b\;\; , \;\;\;\;\;
v_b\equiv\dot{x}_{0}(t)=\frac{2}{\beta}\sinh \beta \; \sin \alpha
\label{omegav}
\end{equation}
where $-\pi \leq \alpha \leq \pi$ and $0 < \beta < \infty$. The
equation (\ref{Salerno}) has the following conserved quantities,
namely the Hamiltonian ${H}$ and the norm ${N}$:
\begin{eqnarray}
{H}&=& -\sum_{n}(\Phi_{n}\overline{\Phi}_{n+1} +
\overline{\Phi}_{n}\Phi_{n+1}) -2\frac{\nu}{\mu}\sum_{n}
|\Phi_n|^2 +2\frac{\nu}{\mu^2}\sum_{n}\ln(1+\mu|\Phi_n|^2)
\label{Ham}
\\
{N} &=& \frac{1}{\mu}\sum_{n}\ln(1+\mu|\Phi_n|^2)
\label{Norm}
\end{eqnarray}
where $\overline{\Phi}_{n}$ denotes the complex conjugate of
$\Phi_{n}$. In what follows we will fix the value $\gamma = 2$ in eq. 
(\ref{DNLS}) and $\mu + \nu = 1$ in eq. (\ref{Salerno}), 
as usual.

Perturbative inverse scattering transform \cite{Vakhnenko}, as well
as collective coordinate methods \cite{Claude,Cai,MackaySep}, have 
been used to study moving breathers of the Salerno equation 
(\ref{Salerno}) near the integrable A-L limit ($\nu\simeq0$). 
The numerical procedure that we explain below has the advantage 
of being unbiased and not restricted to small values of the non-integrability 
parameter $\nu$, at the expense of restricting attention to those 
solutions (\ref{A-Lbreather}) which are {\em resonant}, meaning that 
the two breather time scales are commensurate $2\pi v_b/\omega_b =p/q$ 
(rational time scales ratio). A resonant ($p/q$) moving breather 
$\hat{\Phi}_n(t)$ is 
numerically represented as a fixed point of the map 
${M} = {L}^{p} {T}^{q}$, where 
${L}$ is the lattice translation operator 
${L}(\{\Phi_n(t)\})=\{\Phi_{n+1}(t)\}$, and ${T}$ is the 
$T_b$-evolution map ($T_b = 2\pi /\omega_b$), 
${T}(\{\Phi_n(t)\})=\{\Phi_{n}(t+T_b)\}$; explicitly 
\begin{equation}
\hat{\Phi}_n(t) = \hat{\Phi}_{n+p}(t+qT_b) \;\;\;\;\;\mbox{ for all }\;n
\label{fixed point}
\end{equation}

Let us briefly present the numerical method. The implicit function
theorem 
\cite{Ledermann} ensures a unique 
continuation of a fixed point solution of ${M}$ for parameter 
($\nu$) variations, provided the Jacobian matrix $J=D({M} - I)$ 
is invertible: with this proviso the Newton method \cite{Marin} is an efficient 
numerical algorithm to find the uniquely continued fixed point. 
In other words, continuation from a resonant A-L breather along
the Salerno model is possible if one restricts the Jacobian matrix $J$ to the
subspace orthogonal to its center (null) subspace. The center subspace turns 
out to be spanned by two continuous symmetries of the Salerno model, namely, 
{\em time translation} and {\em gauge} (uniform phase rotation) invariances. 
Using Singular Value Decomposition (SVD) techniques \cite{NumRec}, one then obtains 
numerically
accurate continued resonant moving breathers along the Salerno model until
conditions for continuation cease to hold. A (SVD)-regularized Newton
algorithm was already used by Cretegny and Aubry in
\cite{CretegnyAubry} to refine moving
breathers of Klein-Gordon lattices with Morse potentials obtained by
other means. From the methodological side what is novel here is the 
systematic use of it in order to obtain the family of moving
Schr\"odinger breathers of the NLS lattice (\ref{Salerno}), for 
different values of $2\pi v_{b}/\omega_{b}=0,\; 1/2,\; 3/4,\; 1,...$ 
and a fine grid of frequency values $\omega_{b}$ and the
nonintegrability parameter $\nu$.

\section{Mobile Discrete Breathers}
\label{s.mdb}
Let first start with a few remarks on immobile breathers ($p=0$). Some
nonintegrable issues, that affect mobile solutions, can be shown
continuing the immobile ones along the Salerno Model ($\nu = 0$,{\ldots}, $1$). 
First we remark that the uniquely continued solution of standard 
DNLS ($\nu = 1$) is equal to the pinned discrete breather uniquely continued 
from the anticontinuous limit ($\gamma \rightarrow \infty$) 
\cite{MackayAubry}. 
Second, only inmobile breathers which are centered either 
at a site ($n$) or at a bond  
($n\pm1/2$) persist; this is due to the emergence of Peierls-Nabarro
barriers away from integrability ($\nu\not=0$), 
a well-known result of collective variable theory \cite{Cai,MackaySep}. 
The breather centered at a site is stable while the one 
centered at a bond is unstable \cite{Cai,note}; 
its energy difference is the Peierls-Nabarro barrier. 
This energy difference acts as a barrier to mobile breathers 
for travelling along the lattice; the numerical computations of this
barrier nicely fit with collective variable predictions.

\begin{figure}[!tbh]
\begin{center}
\begin{tabular}{cc}
(a)
\resizebox{5.cm}{!}{%
\includegraphics[angle=-90]{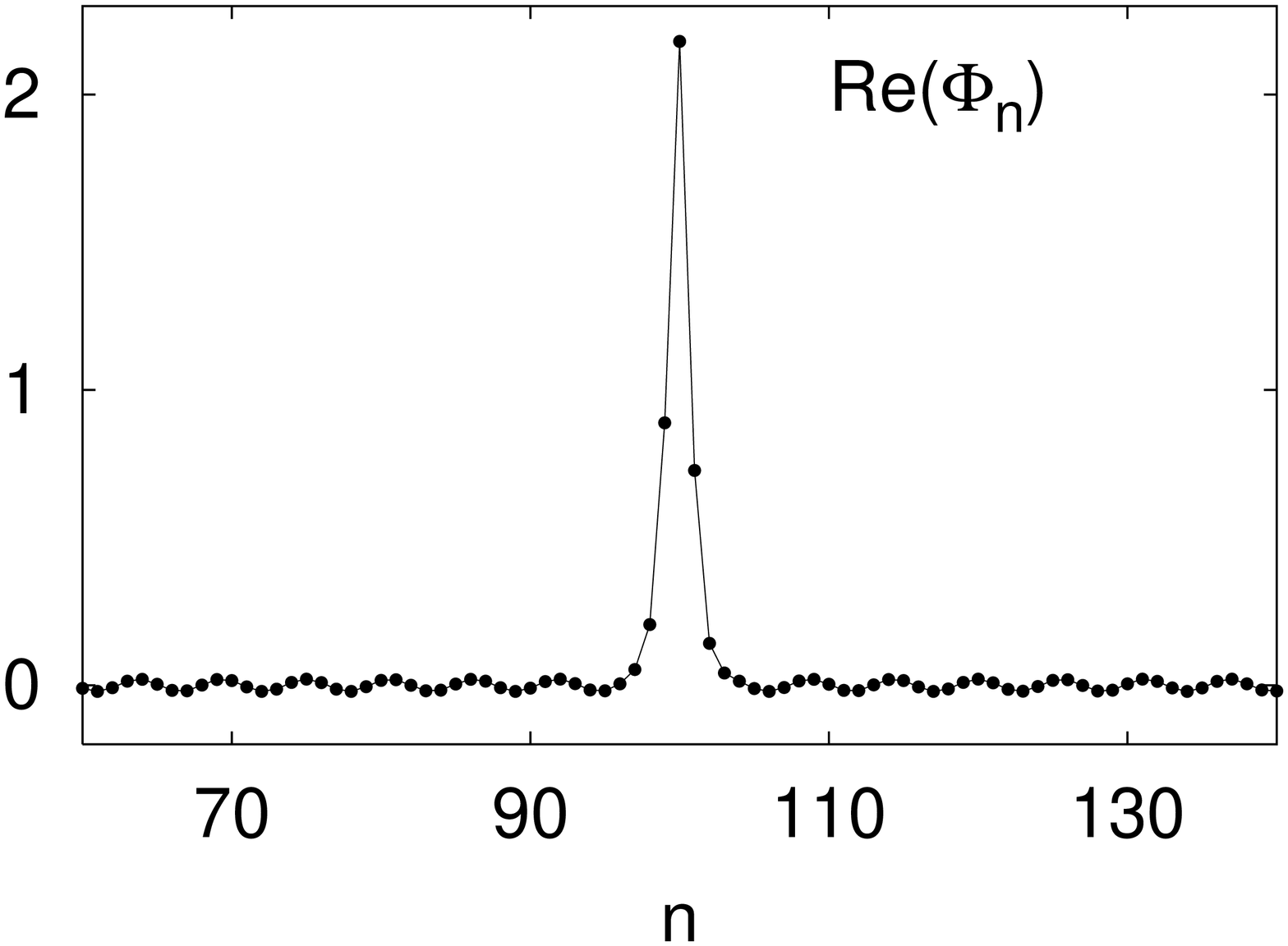}
}
&
(b)
\resizebox{5.cm}{!}{%
\includegraphics[angle=-90]{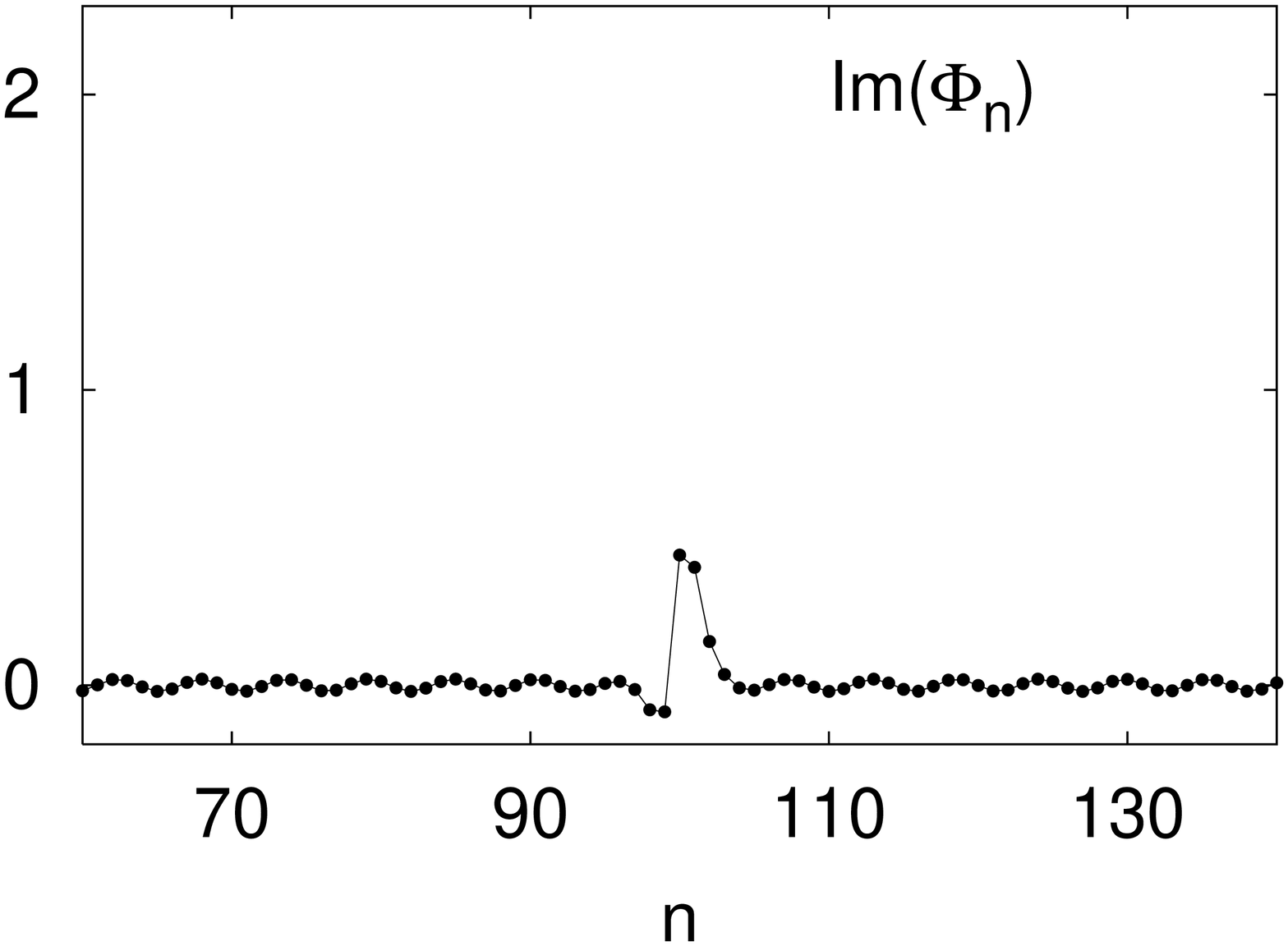}
}
\\
(c)
\resizebox{5.cm}{!}{%
\includegraphics[angle=-90]{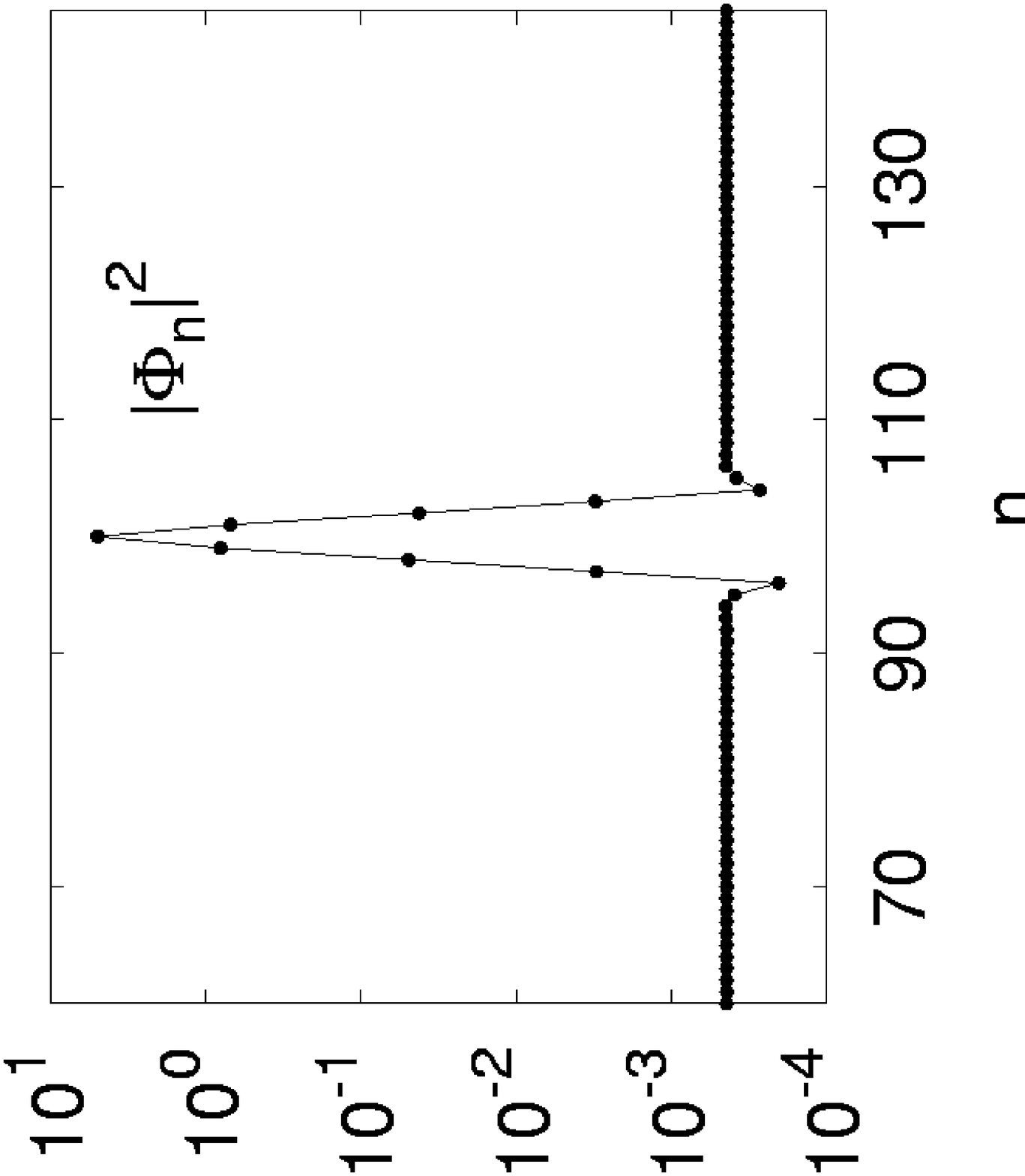}
}
&
(d)
\resizebox{5.3cm}{!}{%
\includegraphics[angle=-90]{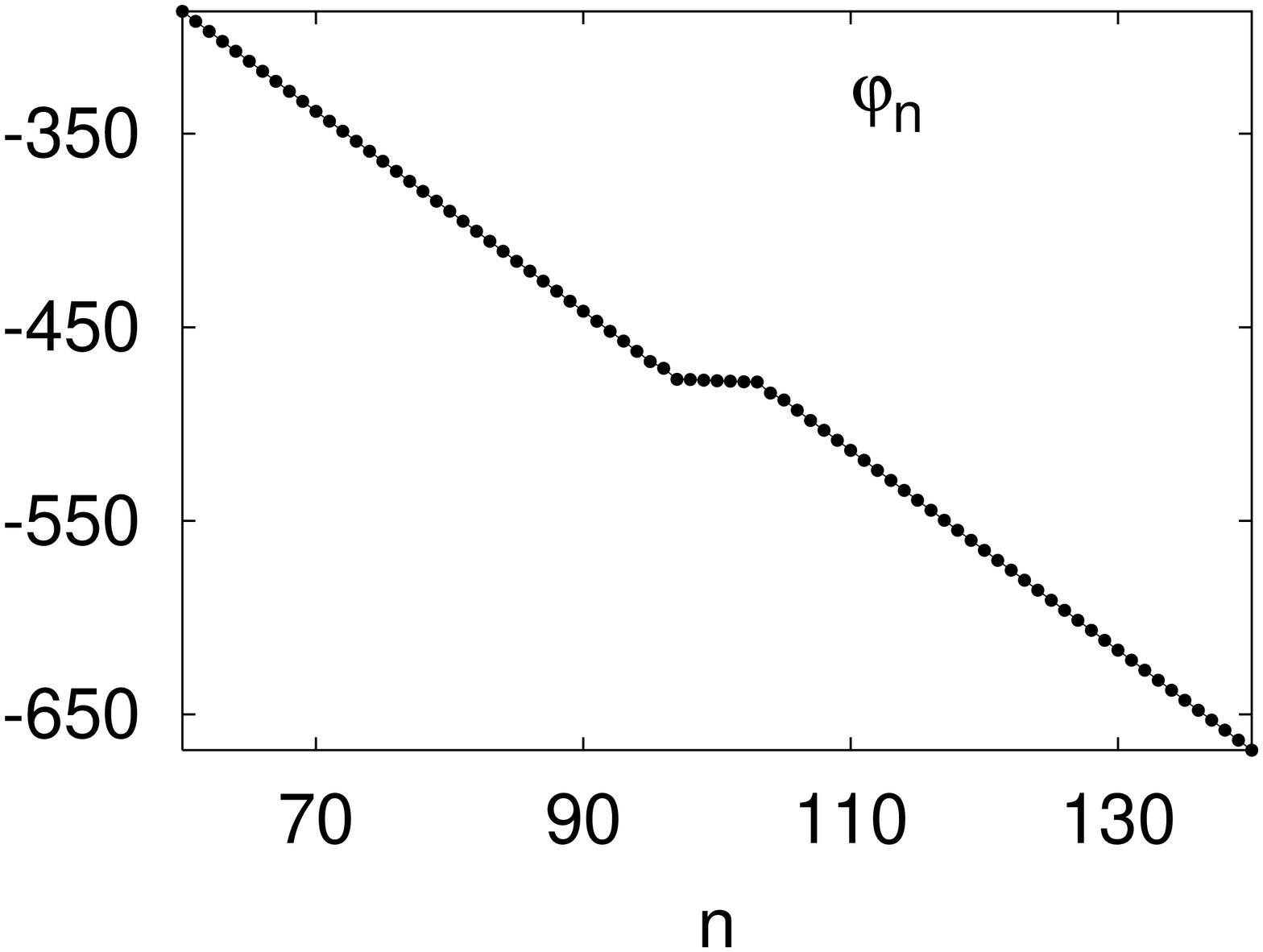}
}
\end{tabular}
\end{center}
\caption{
Instantaneous profile of a $1/1$ breather with $\omega_b=4.909$ and $v_b=0.7813$. 
Real part {\bf (a)}, imaginary part {\bf (b)}, modulus {\bf (c)} and
phase {\bf (d)}. 
The nonintegrability parameter in (\ref{Salerno}) is
$\nu=0.05$.}
\label{fig:solution}
\end{figure}

Our main interest, however, focusses on mobile solutions, 
{\em i.e.} $p \neq 0$. How are Peierls-Nabarro barriers to mobility 
overcome by the fixed point solution?
Our results show clearly that the uniquely 
continued $p/q$-resonant fixed point for 
$\nu \neq 0$ is spatially asymptotic to an {\em extended background}, whose 
amplitude increases from zero (at $\nu =0$) with 
increasing non-integrability $\nu$, superposed to 
the moving (A-L)-like core, see figure 
(\ref{fig:solution}). In order to reveal the structure 
of this extended background, we have to pay attention 
to spatially extended solutions of the Salerno model.

\begin{figure}[!tbh]
\begin{center}
\begin{tabular}{cc}
(a)
\resizebox{6.cm}{!}{%
\includegraphics[angle=-90]{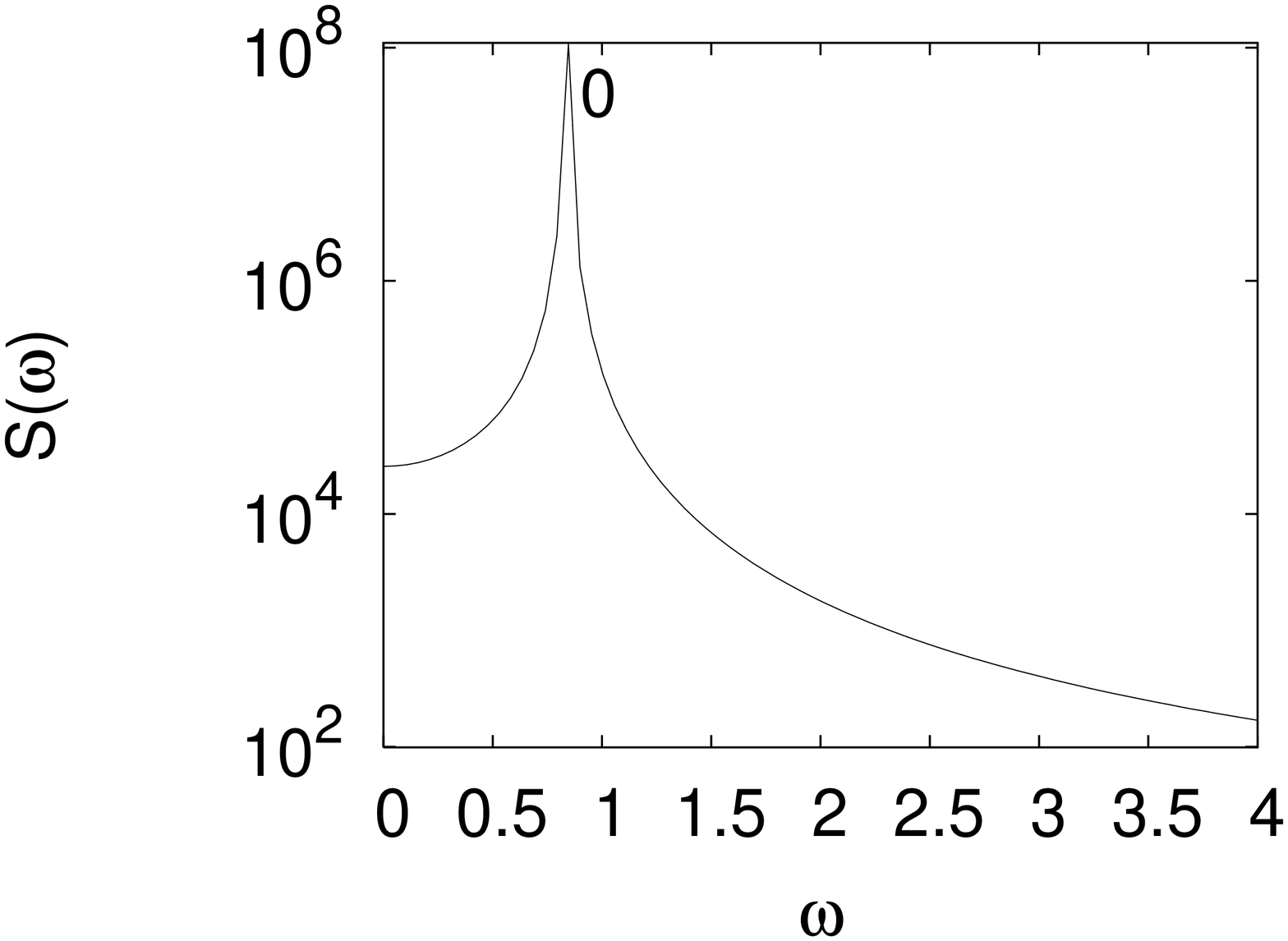}
}
&
(b)
\resizebox{6.cm}{!}{%
\includegraphics[angle=-90]{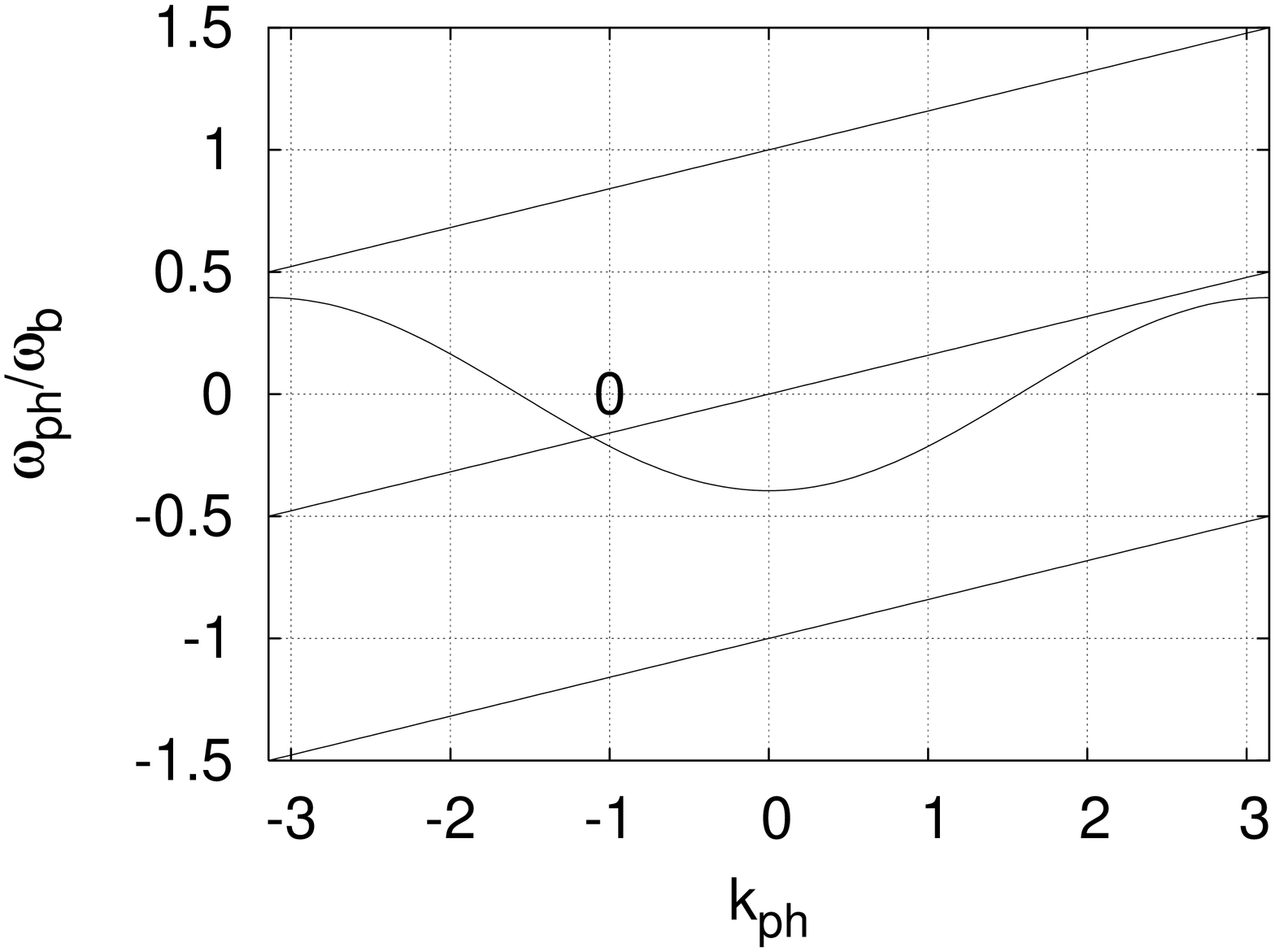}
}
\\
(c)
\resizebox{6.cm}{!}{%
\includegraphics[angle=-90]{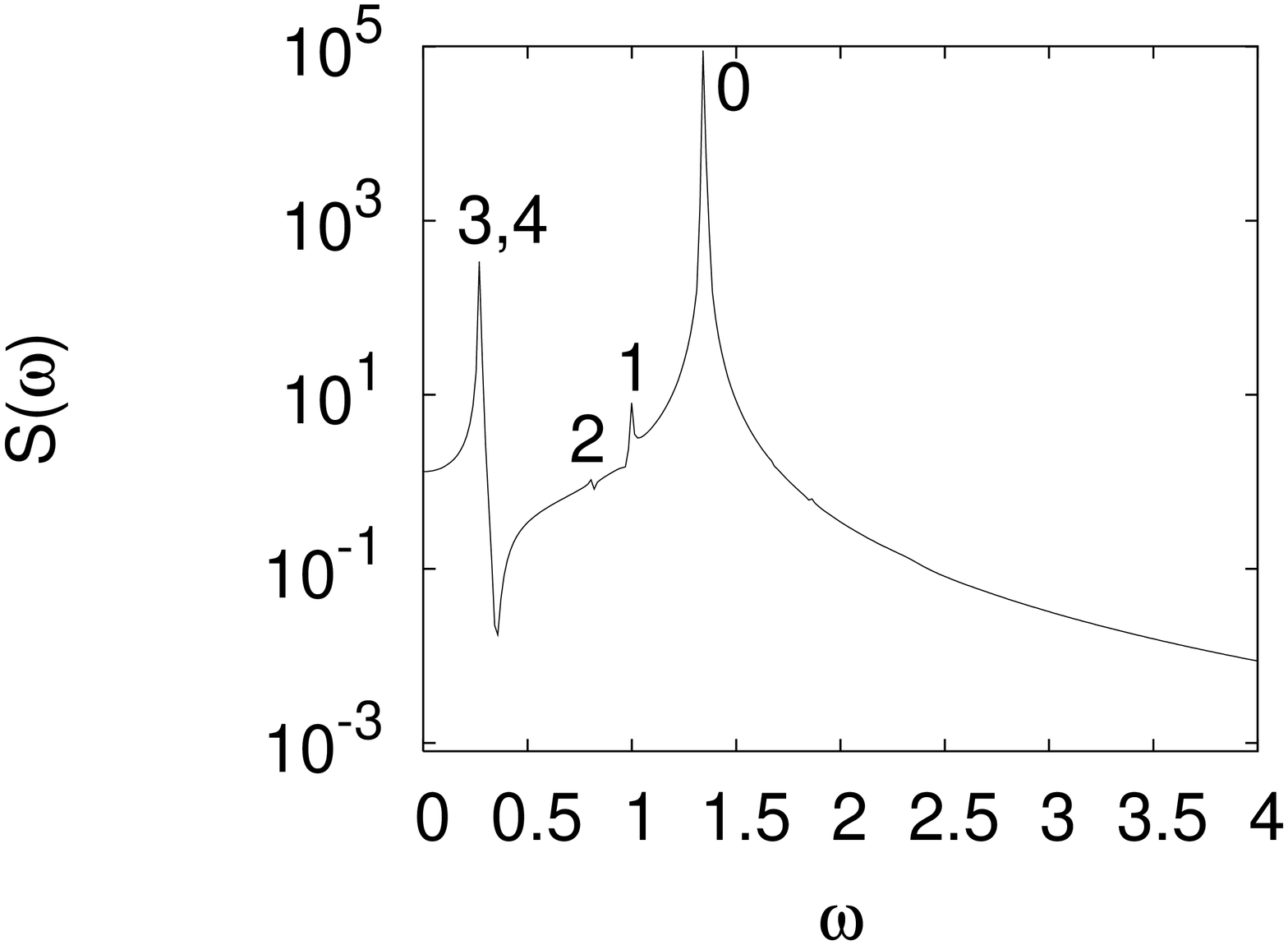}
}
&
(d)
\resizebox{6.cm}{!}{%
\includegraphics[angle=-90]{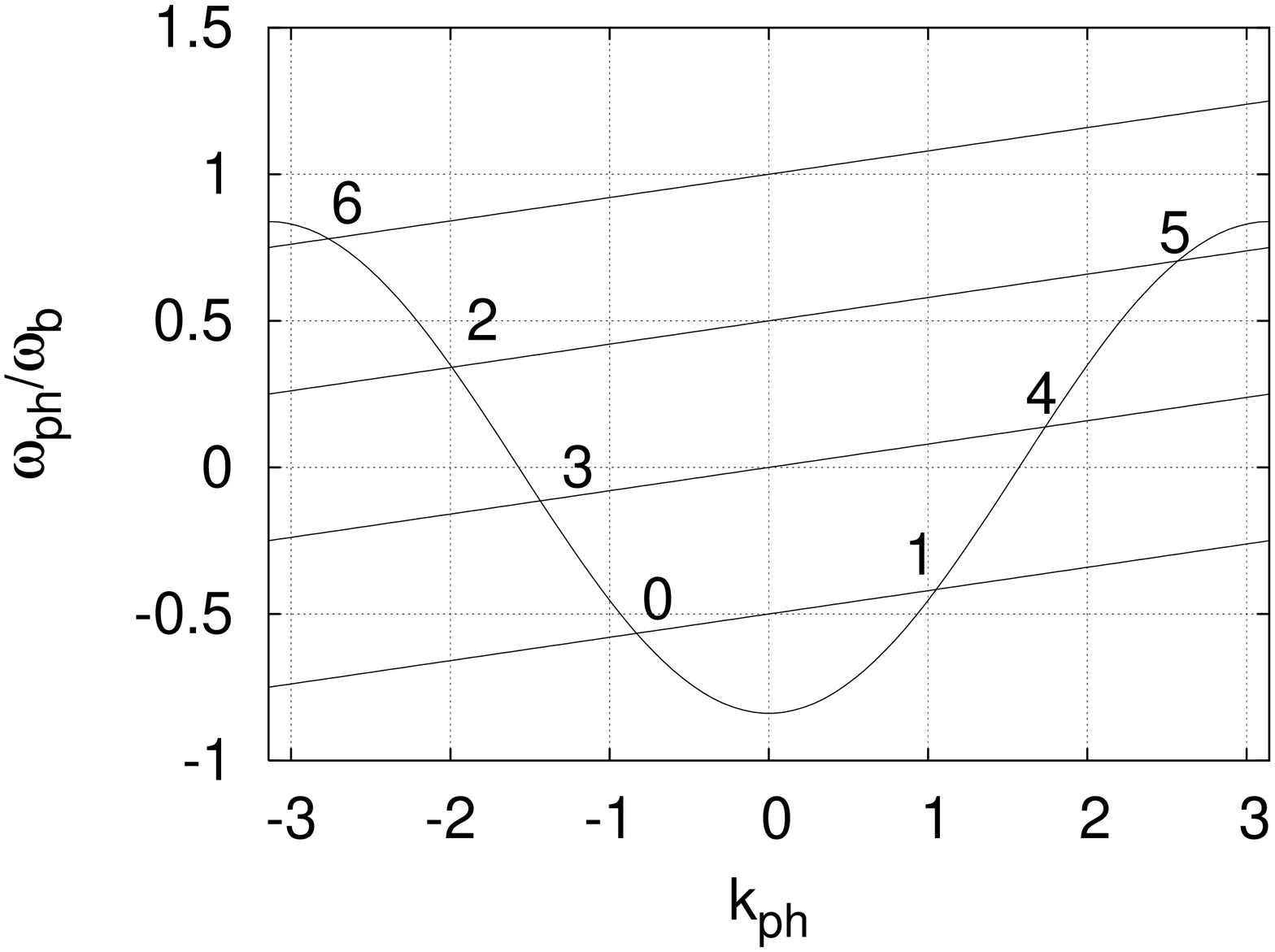}
}
\end{tabular}
\end{center}
\caption{Power Spectrum.
{\bf (a)} $S(\omega)$ for the background of a $1/1$ breather with
$\omega_b=5.057$ 
at $\nu=0.05$. In {\bf (b)} formula (\ref{resonant}) gives the contribution
of a unique phonon  ($j=0$) with which agrees (eq. (\ref{dispersion})) with
results given by $S(\omega)$. 
{\bf (c)} $S(\omega)$ for the background of a $1/2$ breather with
$\omega_b=2.3842$ at $\nu=1.00$. 
In {\bf (d)} formula (\ref{resonant}) gives the contribution
of seven phonons ($j=0,..,6$), but only five of them ($0-4$) are visible on
$S(\omega)$. Note that the amplitudes $|A_{j}|$
differ by orders of magnitude.
}
\label{fig:FFT}
\end{figure}

The Salerno equation (\ref{Salerno}) admits extended solutions of the 
plane wave form, 
$\Phi_n(t) = A \exp[{\mbox i}(k n-\omega t)]$, 
provided the following (nonlinear) dispersion 
relation holds:
\begin{equation}
\omega = -2 [1+(1-\nu)|A|^{2}]\cos k - 2\nu |A|^{2}
\label{dispersion}
\end{equation}
A $p/q$-resonant plane wave satisfies $\Phi_n(t) = \Phi_{n+p}(t+qT_b)$, 
and therefore, for a $p/q$-resonant plane wave the following condition also 
holds: 
\begin{equation}
\frac{\omega}{\omega_b} = \frac{1}{q}\left(\frac{p}{2\pi} k -m  \right)
\label{resonant}
\end{equation}
$m$ being any integer. Equations (\ref{dispersion}) and (\ref{resonant}) 
can be solved for $k$ and one obtains a finite number of branches 
$k_j(|A|)$ of $p/q$-resonant wavenumbers in the first Brillouin zone, 
$-\pi \leq k_j \leq \pi$. 
The simplest case of a unique branch for fixed $\nu$ (as well as 
$\omega_b$ and $p/q$) and $A$ small, is represented in a) and b) of figure (\ref{fig:FFT}). 
For example, for $A$ small and $\omega_b > 4$, for any 
value of $0<\nu<1$ and $A$, there is a unique $1/1$-resonant 
wavenumber branch $k_0(\nu, A)$. 

For the general situation where several branches 
$k_j$ ($j=0, {\ldots},\;s-1 $) of resonant plane waves 
solve (\ref{dispersion}) and (\ref{resonant}), the power 
spectrum of a background site $n$, 
$S(\omega)={\mid\int_{-\infty}^{\infty}{\Re(\Phi_{n}(t))\exp\lbrack{\mbox
i}\omega t\rbrack dt}\mid^2}$, reveals $s$ peaks at the values 
$\omega_j$ corresponding to the resonant 
wavevector branches. The background is, up to numerical accuracy, a
linear superposition of $p/q$-resonant plane waves, namely
\begin{displaymath}
\sum_{j=0}^{s-1} A_j \exp[{\mbox i}(k_j n-\omega_j t)]
\end{displaymath}
The amplitudes $A_j$ 
differ typically orders of magnitude, {\em i.e.} 
$|A_0| \gg |A_1| \gg |A_2| {\ldots} $, so that only a few 
frequencies are dominant, for most practical purposes. 
One would speak of localization in $k$-space to describe 
the extended background of the $p/q$-resonant fixed point. 
Once the values of $\omega_b$, 
$v_b$ and $\nu$ are given, the ``selection rule'' provided 
by equations (\ref{dispersion}) and (\ref{resonant}), 
does not determine directly the resonant wavenumbers 
$k_j$, but only branches $k_j(A)$. This reflects the 
inherent nonlinearity of the NLS lattice, wherefrom 
the frequency of the plane wave depends on both 
wavenumber and amplitude in equation (\ref{dispersion}). 
Along the parametric continuation path the fixed point ``adjusts'' the 
planewave content ($k_j$) of the background, so that it remains 
$p/q$-resonant under the changes in the amplitudes of the background plane waves
($A_j$). 

\section{Background relevance to Mobility}
\label{s.bckg}
\begin{figure}[!tbh]
\begin{center}
\begin{tabular}{cc}
{(a)}
\resizebox{8cm}{!}{
\includegraphics[angle=-90]{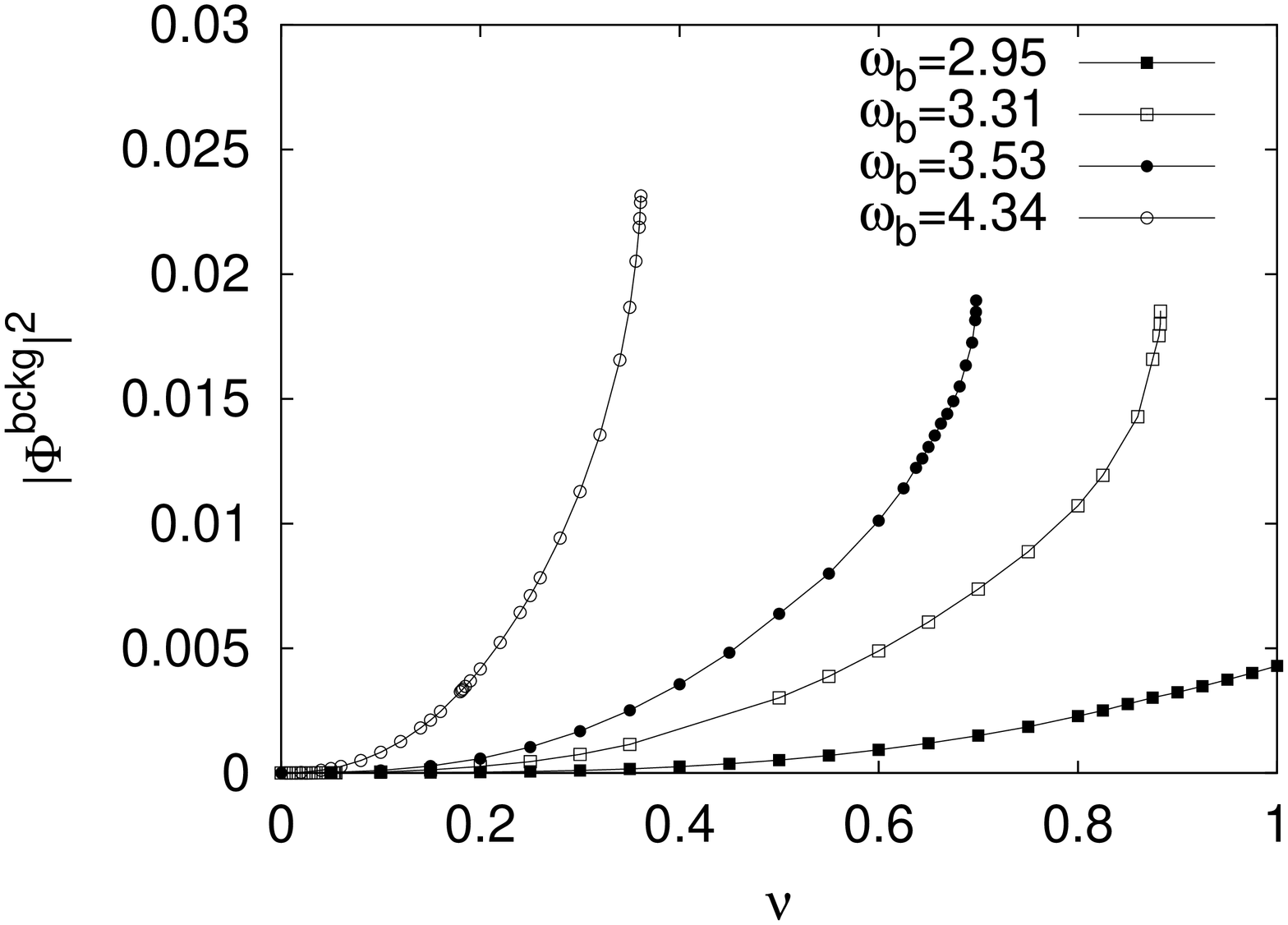}
}
\\
{(b)}
\resizebox{9cm}{!}{
\includegraphics[angle=-90]{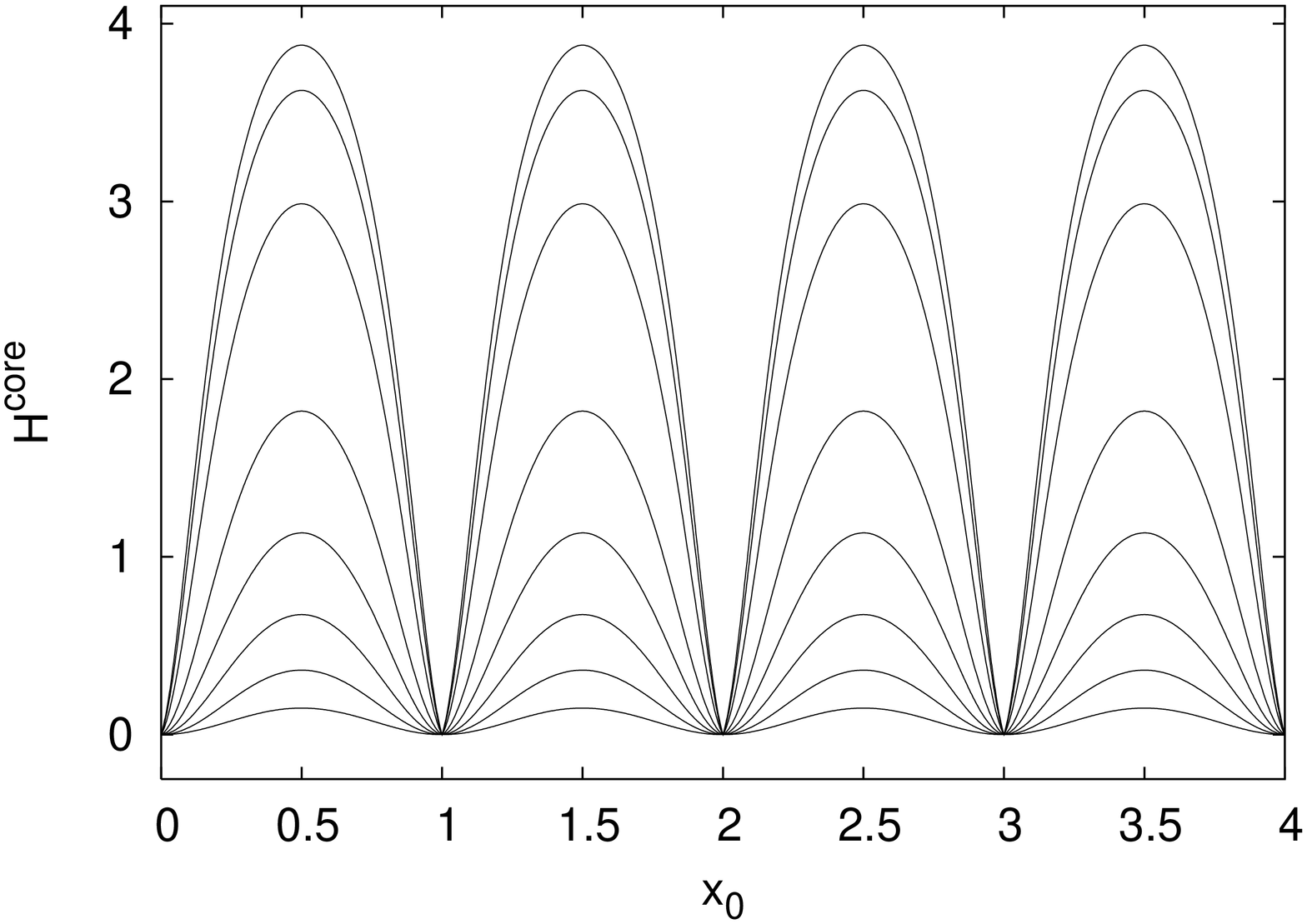}
}
\end{tabular}
\end{center}
\caption{
{\bf (a)} Background amplitude of different $1/1$ resonant breathers
as a function of the nonintegrability $\nu$. The amplitudes are zero
in the A-L integrable limit ($\nu=0$) and have a monotone increasing
behavior with $\nu$. The value $\nu=1$ corresponds to the standard
DNLS equation. 
{\bf (b)} Plot of $H^{core}$ of a $1/1$ resonant breather
with $\omega_{b}=5.056$ as a function of the 
localization center $x_{0}$ for different values of $\nu$ ($0.04$, $0.08$, $0.12$, $0.16$,
$0.20$, $0.24$, $0.25$ and $0.2512$ (end of the continuation)).  
The amplitude of the oscillation of $H^{core}$ grows with $\nu$. (The
minimum value of $H^{core}$ has been set to zero in order to compare the 
differents functions.)}
\label{fig:bckg_Ham}
\end{figure}

Along the Newton continuation path to the standard DNLS equation the background
amplitudes have a monotone increasing behavior with $\nu$, see figure 
(\ref{fig:bckg_Ham}.a). High frequency solutions cannot be continued up
to that limit, and the continuation stops for values of $\nu\;<\;1$. 
This result correlates well with the collective variable (particle perspective) 
predictions \cite{Claude,Kundu} where the non-persistence of
travelling solutions is related to the growth of the
Peierls-Nabarro barrier. In this respect, one observes a sudden
increase in the background amplitude near the continuation
border. This result reinforces the interpretation of the background as
an energy support to the core for surpassing the (nonintegrable)
Peierls-Nabarro barriers to mobility, and so its unavoidable presence for the
existence of mobile breathers in the absence of integrability.  

The role of the background in the localized core
mobility can be analysed as follows. 
As the solution is unambiguosly found to be
$\hat{\Phi}=\hat{\Phi}^{core}+\hat{\Phi}^{bckg}$, the energy
${H}$, equation  (\ref{Ham}), of a mobile breather can be written as
\begin{equation}
{H}={H}[\hat{\Phi}^{core}]+{H}[\hat{\Phi}^{bckg}]+{H}^{int}
\label{ham_dec}
\end{equation}
where ${H}^{int}$ is the interaction energy, {\em i.e.} the
crossed terms of $\hat{\Phi}^{core}$ and $\hat{\Phi}^{bckg}$ 
in the Hamiltonian. In the simplest case in which the background 
has a single resonant plane wave, its energy is a constant of 
motion (along with the total energy), so one obtains
\begin{equation}
\frac{\partial {H}[\hat{\Phi}^{core}]}{\partial t}=
-\frac{\partial {H}^{int}}{\partial t}
\label{balance}
\end{equation}
{\em i.e.} the variations of the core energy along the motion are
balanced by the variations in the core-background interaction
energy. Equation (\ref{balance}) dictates the dynamics of any
(eventual) effective (collective) variables intended to describe the
mobile core in a particle-like description of the breather.

One can compute the core energy variations directly from the numerical
integration of a solution by substracting (at each time step) the
background from it. In figure (\ref{fig:bckg_Ham}.b) we plot the evolution of
the core energy as a function of the core localization center,
$x_{0}$, defined by using the norm ${N}$ (equation (\ref{Norm})) of
the Salerno model (\ref{Salerno}) as
\begin{equation}
x_{0}=\frac{\sum_{n}{n\ln(1+\mu|\Phi_{n}^{core}|^{2})}}{\mu{N}}.
\label{loc_center}
\end{equation}
One observes that the core has extracted the
maximum energy from the interaction term when the core passes over
$x_{0}=n\pm 1/2$ (maxima of the PN barrier) and has given it back to
the interaction term at $x_{0}=n$ (minima of the PN barrier). The
bigger the PN barrier, the larger the interaction term (directly
proportional to background amplitude) is. This
result illustrate the role of the resonant background on the core
mobility and the interpetation of its amplitude increase with $\nu$.
{\em The increase of nonintegrability, and the subsequent growth of the
PN barrier, demands aditional support of energy from the interaction
term, which is achieved by an increase of the background amplitude}. 

\section{Conclusions}
\label{s.conc}
We have used a (SVD)-regularized Newton algorithm  to
continue mobile discrete breathers in the Salerno model
from the integrable Ablowitz-Ladik limit. Our results indicate that, away from 
integrability, a description of these solutions based exclusively on
localized (collective) variables is incomplete. The
solutions are composed by a localized core and a linear superposition
of plane waves, the background, whose amplitudes differ orders of
magnitude. The background plays an important role in the translational
motion of the localized core. Exact mobile localization only exist
over finely tuned extended states of the nonlinear lattice. Mobile
``pure'' ({\em i.e.} zero background) localization must be regarded as
very exceptional (integrability).

\section{Acknowledgments}

Financial support came from MCyT 
(BFM2002-00113, I3P-BPD2002-1) and LOCNET HPRN-CT-1999-00163. 
The authors acknowledge A.R. Bishop for arising our interest in DNLS, 
P. Kevrekidis, J.L. Garc\'{\i}a-Palacios, S. Flach, R.S. Mackay,
M. Peyrard and G.P. Tsironis for discussions on our numerics 
and sharing his intuitions.


\begin{thebibliography}{0}
\bibitem{note}
  That is the case for unstaggered ($\alpha =0$)-continued 
stationary breathers from A-L. The staggered breathers ($\alpha =\pi$) 
have its stability reversed, as shown in \cite{Cai}.
\end{thebibliography}
\end{document}